\documentclass[%
 reprint,
superscriptaddress,
nofootinbib,
 amsmath,amssymb,
 aps,
]{revtex4-2}

\usepackage{graphicx}
\usepackage{dcolumn}
\usepackage{bm}
\usepackage{hyperref}
\hypersetup{colorlinks=true, linkcolor=blue!80!black, urlcolor=blue!80!black, citecolor=blue!80!black}

\usepackage{floatrow}

\usepackage{xcolor,marginnote}
\usepackage{etoolbox}
\makeatletter
\patchcmd{\@mn@@@marginnote}{\begingroup}{\begingroup\@twosidefalse}{}{\fail}
\makeatother
\usepackage{braket}
\usepackage[nameinlink]{cleveref}

\newcommand{\maximize}{\operatornamewithlimits{maximize}}

\reversemarginpar

\begin{document}

\preprint{APS/123-QED}

\title{Efficient Frequency Allocation for Superconducting Quantum Processors Using Improved Optimization Techniques}

\author{Zewen Zhang}
\affiliation{Mathematics and Computer Science Division, Argonne National Laboratory, Lemont, Illinois 60439, USA}
\affiliation{Department of Physics and Astronomy, Rice University, Houston, Texas 77005, USA}
\author{Pranav Gokhale}
\affiliation{Infleqtion, Chicago, Illinois 60604, USA}
\author{Jeffrey M. Larson}%
\affiliation{Mathematics and Computer Science Division, Argonne National Laboratory, Lemont, Illinois 60439, USA}

\date{\today}

\begin{abstract}
Building on previous research on frequency allocation optimization 
for superconducting circuit quantum processors, this
work incorporates several new techniques to improve overall solution
quality. New features include tightening constraints, imposing edgewise
differences, including edge orientation in the optimization, and integrating
multimodule designs with various boundary conditions. These enhancements allow
for greater flexibility in processor design by eliminating the need for
handpicked orientations. We support the efficient assembly of large processors
with dense connectivity by choosing the best boundary conditions. Examples
demonstrate that, at low computational cost, the new optimization approach
finds a frequency configuration for a square chip with over 1,000 qubits and
over 10\% yield at much larger dispersion levels than required by previous
approaches. 
\end{abstract}

\maketitle


\section{\label{sec:intro}Introduction}
Superconducting (SC) quantum processors have led to significant progress in
numerous domains, including demonstrations of quantum
advantage~\cite{arute2019quantum,kim2023evidence}, quantum error
correction~\cite{andersen2020repeated,sundaresan2023demonstrating}, variational
methods~\cite{harrigan2021quantum,maciejewski2023design,zhang2023simulating,dupont2024quantum},
and analog simulations~\cite{orell2019probing,hung2021quantum}. Among competing
quantum architectures, fixed-frequency SC devices are a leading technology due
to their high coherence, relatively simple control, and ability to scale up to
hundreds of qubits in state-of-the-art
devices~\cite{yu2023simulating,kim2023evidence}. However, the need for
addressability, which requires large detunings among neighboring qubits, and
controllability, which is set by the entangling protocol, imposes constraints
on selecting local frequencies for each
qubit~\cite{brink2018device,morvan2022optimizing,zhao2023mitigation}; these constraints present challenges in
SC chip manufacturing and hinder the scalability and connectivity of these
devices. Therefore, solving frequency allocation problems is critical to the
success of fixed-frequency SC devices.

Moreover, recent developments in quantum error correction have further
increased the potential for frequency crowding, heightening the need
for improved techniques for avoiding frequency collision. In particular, while
the nearest-neighbor surface code \cite{fowler2012surface} quantum error
correction 
architecture has historically been favored for SC processors,
advances in the study of non-local qLDPC
codes~\cite{breuckmann2021quantum,tillich2013quantum} have placed them at the forefront of emerging roadmaps. We highlight two examples here. One recent work by IBM~\cite{bravyi2024high} has discovered distance-12
qLDPC codes that encode 12 logical qubits into 144 physical qubits---this is over 10 times more space-efficient than the equivalent surface code. However, this code requires degree-6 qubit connectivity, substantially
increasing the frequency collision risk over degree-4 surface code
interaction graphs. While the frequency collision overhead could be mitigated by switching to an SC device with tunable couplers and/or
frequency-tunable qubits, doing so would add significant complexity to control and wiring requirements. 
In a similar spirit, another recent work~\cite{poole2024architecture} demonstrates that IBM's code could be remapped to a single plane that
permits two-qubit gates across qubits up to 7.22 units away, that is, (6,4) units away in taxicab geometry. While this approach is favored from the perspective of having shorter long-range connections, it would lead to even greater chances of frequency collisions because of the relative geometric locality of the frequency crowding.
Thus, while non-local qLDPC codes offer a promising path forward for more efficient quantum error correction, they also amplify challenges associated with frequency collisions.

Another prevailing trend that induces even greater conditions on frequency
requirements is an 
increased interest in qutrit and qudit systems in quantum applications. As a
baseline, many two-qubit gates in SC systems in fact use
qutrit states to perform entangling operations. In addition, IBM now
routinely performs qubit readout via excited state promotion, a technique that
excites $\ket{1}$ to $\ket{2}$ prior to measurement
\cite{jurcevic2021demonstration}. Moreover, algorithmic research
\cite{gokhale2019asymptotic} has demonstrated that computing with qutrits can
achieve an exponential advantage in runtime over qubits. Because of these
developments, quantum hardware designers are motivated to support qudit
control~\cite{campbell2023superstaq}; however, this shift further increases the risk of frequency
collisions, as additional relevant frequencies (e.g., $\ket{1}
\rightarrow \ket{2}$) must also be avoided.

When designing an SC device, the individual frequency $f_i$ of each qubit $i$
is determined by the critical current of the Josephson junction  and the
circuit's capacitance. As with any manufacturing process, frequencies are often
fabricated with a certain
degree of uncertainty, which can lead to constraint violations between
neighboring qubits (``frequency collision") and, consequently, reduce the success
ratio of producing collision-free processors (``yield") during manufacturing. 
Various optimization methods have been implemented to address frequency
allocation in SC devices, aiming to reduce the probability of frequency
collisions and improve yield when manufacturing chips, while balancing
computational resources required for optimization strategies. 
Even with novel optimization methods, scaling SC quantum
processors with desired connectivity to hundreds of qubits is challenging and often
results in an exponentially low yield. Significant progress in solving
frequency allocation problems has been made using numerical tools such as
mixed-integer-programming-based optimization~\cite{morvan2022optimizing},
multichip module approaches~\cite{smith2022scaling}, physics-inspired
frameworks~\cite{zhang2024qplacer}, Floquet analysis~\cite{heya2024floquet}, and connectivity
optimization~\cite{lin2022domain,yang2023superconducting}. 
Improvements in SC qubit
fabrication have also helped mitigate frequency
collisions~\cite{kim2022effects,zhang2022high,balaji2024electron}. Optimization
challenges persist, however, hindering the production of ideal (large) devices with 
good connectivity, such as the native devices required for square grid
structures for the surface code~\cite{fowler2012surface} or the degree-6
connectivity needed for 
advanced codes~\cite{bravyi2024high}.

In this work we develop several optimization techniques that build on the
mixed-integer-programming-based approach in Ref.~\cite{morvan2022optimizing} to
significantly improve device yield compared with previous methods. First, by
tightening certain constraints in specific allocation problems, we achieve
substantial yield improvements. Second, we address the often overlooked issue
of qubit orientation (i.e., determining control and target qubits for each
edge) and introduce it as a decision in the optimization process. We also
consider adding edgewise differences to our optimization model to avoid some
special symmetric solutions allowed by the topology and the objective function,
which often lead to low yield in practice. Third, we enhance multichip
module design~\cite{smith2022scaling} by incorporating various boundary
conditions in assembling small modules, allowing the methods to scale to larger
problems. With these improved optimization techniques, we can compute frequency
configurations for $32\times32$ square grid chips with over 10\% yield at a threshold
dispersion of 6.5 MHz, which is around 30\% better than the previous
results~\cite{morvan2022optimizing}.

We now give an outline of the paper. \Cref{sec:formulation} discusses three
significant optimization approaches for improving the solution quality to the
frequency allocation problem. \Cref{sec:results} presents results for how these
approaches improve solution quality. \Cref{sec:outlook} concludes with an
outlook on future approaches for addressing frequency allocation problems.

\section{Frequency allocation and optimization formulation\label{sec:formulation}}

The fabrication process of SC
qubits introduces a stochastic frequency
dispersion $\sigma$ to the value of each desired $f_i$, which can result in unwanted
crosstalk~\cite{zhang2024qplacer}. This dispersion, primarily arising during
the fabrication of Josephson
junctions~\cite{berke2022transmon,osman2023mitigation}, maintains a value of around 10 MHz even in the simplest SC qubit structure, the transmon platforms~\cite{zhang2022high,hertzberg2021laser}.
If exact frequencies could be fabricated (i.e., if $\sigma$ was zero), the frequency allocation problem would be greatly simplified.
Here we  use
numerical optimization to determine frequencies $f_i$ that account for this
dispersion. Frequency collisions can arise for a variety of physical reasons,
each of which  can be modeled as a constraint in our numerical optimization
model. 
We consider the same collisions studied in
Refs.~\cite{morvan2022optimizing,hertzberg2021laser}; their corresponding constraints are listed for completeness
in \Cref{table:constraint}. There are nine families of constraints for SC 
processors with cross-resonance~\cite{rigetti2010fully} (CR) entangling gates, which 
can be sorted in three categories: Addressability (\textbf{A1} and
\textbf{A2}), Entanglement (\textbf{C1}, \textbf{D1}, \textbf{E1}, and
\textbf{E2}), and Spectator (\textbf{S1}, \textbf{S2}, and
\textbf{T1}) constraints~\cite{brink2018device}.

A frequency allocation problem is commonly modeled as occurring on a graph. In
this graph the set of vertices ($V$) represent qubits, and the set of edges ($E$)
represent entangling gates connecting two qubits. A real value $f_i$ is assigned to vertex $v_i$ in $V$ to represent the desired frequency for the $i$th qubit.
Within the CR protocol,
orientations must be chosen with specific \textit{control} qubit $\rightarrow$
\textit{target} qubit directions. 
In the notation, besides the undirected edge set $E$, the directed edge set is denoted $\vec{E}$, with $(i, j) \in \vec{E}$, which indicates qubit $i$ is the \textit{control} qubit and qubit $j$ is the \textit{target} qubit. This notation is specifically associated with the entanglement constraints outlined in~\Cref{table:constraint}.  In
~\Cref{sec:optimize_orientation} and beyond, however, as the orientation becomes subject to optimization, no constraints are imposed based on the initial input orientation.

\begin{table}[ht]
\centering
\begin{tabular}{|>{\centering\arraybackslash}m{0.8cm}|>{\centering\arraybackslash}m{3.6cm}|>{\centering\arraybackslash}m{2cm}|>{\centering\arraybackslash}m{1.4cm}|}
\hline
\textbf{Type } & \textbf{Definition} & \textbf{Participants} & \textbf{Bounds} \\
\hline
\textbf{A1}& $\left| f_i - f_j \right| \geq \delta_{A1}$ & $(i, j) \in E$ & 17 MHz \\
\hline
\textbf{A2} & $\left| f_i - f_j - \alpha \right| \geq \delta_{A2}$ & $(i, j) \in E$ & 30 MHz \\
\hline
\textbf{C1} & $f_i + \alpha \leq f_d \leq f_i$ & $(i, j) \in \vec{E}$ & --- \\
\hline
\textbf{E1} & $\left| f_d - f_i \right| \geq \delta_{E1}$ & $(i, j) \in \vec{E}$ & 17 MHz \\
\hline
\textbf{E2} & $\left| f_d - f_i - \alpha \right| \geq \delta_{E2}$ & $(i, j) \in \vec{E}$ & 30 MHz \\
\hline
\textbf{D1} & $\left| f_d - f_i - \alpha/2 \right| \geq \delta_{D1}$ & $(i, j) \in \vec{E}$ & 2 MHz \\
\hline
\textbf{S1} & $\left| f_d - f_k \right| \geq \delta_{S1}$ & $(i, j, k) \in N$ & 17 MHz \\
\hline
\textbf{S2} & $\left| f_d - f_k - \alpha \right| \geq \delta_{S2}$ & $(i, j, k) \in N$ & 25 MHz \\
\hline
\textbf{T1} & $\left| f_d + f_k - 2f_i - \alpha \right| \geq \delta_{T1}$ & $(i, j, k) \in N$ & 17 MHz \\
\hline
\end{tabular}
\caption{\label{table:constraint} Original constraints for directed graphs,
cited from Ref.~\cite{morvan2022optimizing}. 
$(i, j, k) \in N$ when $(i, j)
\in \vec{E}$ and $(j, k) \in E$. $\alpha=-0.35$ GHz is the anharmonicity of qubits. In the CR protocol, for connected qubit pair $i\rightarrow j$, the driving
frequency $f_d$ on the control qubit $i$ is equal to the frequency $f_j$ of the
target qubit $j$~\cite{rigetti2010fully}.}
\end{table}

To ensure that the chosen qubit frequencies do not incur collisions, we
introduce a variable $\delta_n$ for each of the nine types of constraints
(for example, $\delta_{A1}$ for the \textbf{A1} constraints $\left| f_i - f_j
\right| \geq \delta_{A1}$). Merely setting an objective to $\maximize \sum_n
\delta_n$ could result in several values of $\delta_n$ being large (meaning those types of collisions
are unlikely to happen), but other values of $\delta_n$ may be zero (meaning
collisions of that type are likely to occur with finite dispersion). Instead, we
set lower bounds $\bar{\delta}_n$ for each of the $\delta_n$ and then set our
objective to be $\maximize \sum_n (\delta_n-\bar{\delta}_n)$. The values for
each $\bar{\delta}_n$ are given in \Cref{table:constraint}.

Ultimately, we model the frequency allocation problem as a mixed-integer
programming problem; we have made our models freely available\footnote{ source code:
\url{https://github.com/AlvinZewen/SC_Freq_Allo}}. We hope that these open-source
models support
the frequency allocation process in SC devices and inspire other researchers to
contribute to their optimization. The numerical results presented below utilize
these model with CPLEX as the optimization solver.

As we show, the adverse effects of dispersion can be mitigated through proper optimization
strategies. These strategies improve the yield in the fabrication process. 
\Cref{sec:edgewise} introduces our tightened constraints and edgewise-difference constraints;
\Cref{sec:optimize_orientation} introduces the orientation of edges as optimization parameters; and 
\Cref{sec:multichip} explains how we can assemble smaller chips into larger chips with various boundary conditions.

\subsection{Constraint Tightening and Edgewise Differences}\label{sec:edgewise}
For a specific topology (i.e., set of edges $E$), we often observe that frequency collisions 
are more likely to occur for certain types of  constraints. Since the lower bounds of these constraints are set by the
fidelity requirement~\cite{magesan2020effective}, the lower bounds on these types of constraints can be increased for better results. By tightening some constraints---increasing
the relevant $\delta_n$ values before the optimization process---the yield can be
significantly enhanced, as we will demonstrate in \Cref{sec:tighten}. However,
excessively increasing $\delta_n$ causes other types of frequency
collisions, especially when modeling devices with many qubits. From the
optimization perspective, appropriately tightening constraints, especially to
account for fabrication dispersion, can considerably improve the yield.
Therefore, in practice, it is advisable  either to identify the constraints that
are more frequently violated for a given problem and impose stricter bounds on
those or, more broadly, to tighten all constraints by a tolerance
$\epsilon_\text{tol}$, typically set as a multiple of $\sigma$, in order to prevent
collisions that might arise from selectively tightening only certain
constraints. 

\begin{figure}[h]%
\centering
\includegraphics[width=0.99\linewidth]{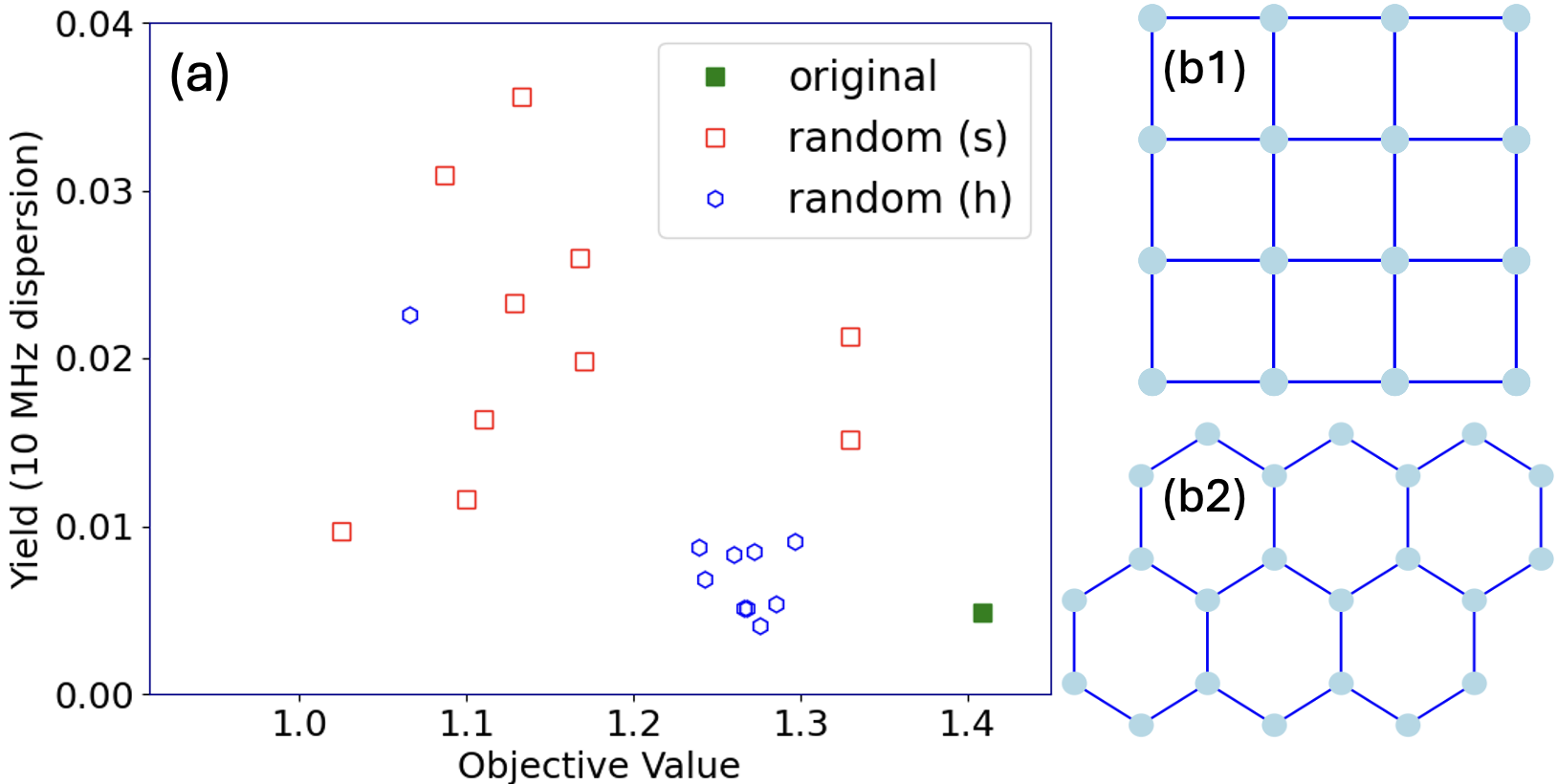}
\caption{Yield at dispersion of 10 MHz for different orientations. (a) 
The optimized objective and yield for $4\times 4$ square (s) grid (b1) and
6-ring hexagon (h) grid (b2). Green solid box: 4$\times$4 square grid using the
orientation in Ref.~\cite{morvan2022optimizing}; red open box: 4$\times$4 square
grid using random orientations; blue open hexagon: 6-ring hexagon grid using
random orientations.} \label{fig:obj}
\end{figure}

As with any mathematical model of a physical system,  a disconnect can exist
between the predicted and actual effect of selecting a set of frequencies. A
higher value of the model objective does not necessarily ensure a
higher yield at certain dispersion levels, as shown in \Cref{fig:obj}. In the figure, yields and objective values of optimized solutions are compared for different fixed orientations. For both square and hexagon grids, 10 randomly picked orientations are preset before optimization for comparison.
We find that much of the disconnect between the model objective and the yield occurs because some solutions have multiple edges that are
exactly at their constraint lower bounds $\bar{\delta}_n$.
For example, consider the optimal
solution on a $4\times4$ grid. Previous optimization methods tend to provide a
near-symmetric solution along one of the diagonals, in which most of the edges on
the boundary are stuck at the lower bound of constraint \textbf{A1} and
eventually give low yields. Merely selecting/requiring that $f_i$ differ for
each vertex in $V$ can still result in a value of $\| f_i - f_j\|$  that is close to or
identical for edges $(i,j) \in E$. 

Such duplication of edge differences may be inevitable for
some optimization models and can be especially problematic when topologies
contain multiple symmetries. In \Cref{fig:obj} we can see  that for given
topologies, the orientation that maximizes the
model objective does not provide the highest yield at a dispersion of 10 MHz. To
address this situation, we impose constraints to ensure there are differences between
the frequencies along the edges in the solution.
For any two edges $k$ and $l$ with no common qubits shared in the graph,
given $f_{k1}$ and $f_{k2}$ ($f_{l1}$ and $f_{l2}$) as the frequencies of the
two qubits attached to edge $k$ ($l$), we impose
\begin{equation}
\label{eq:diff}
    ||f_{k1}-f_{k2}|-|f_{l1}-f_{l2}||<\delta_\text{diff},
\end{equation}
where $\delta_\text{diff}$ is chosen as a small value (empirically $1\sim
5$ MHz).  This rules out solutions where multiple edges are stuck at the lower
bound of a specific constraint. For example, following the imposition of
\Cref{eq:diff}, an optimized solution will not contain more than one edge
positioned at the lower bound of \textbf{A1} or on the left side of 
\textbf{C1}. 
For two edges sharing a qubit, their difference is handled by
the spectator constraint \textbf{S1} in \Cref{table:constraint}. 

\subsection{Optimization of Edge Orientation}\label{sec:optimize_orientation}

Naturally, the optimized frequency configuration and especially the resulting
yield can depend significantly on the chosen orientation of control and target qubits. 
The choice of orientation is rarely discussed in earlier research. But as we observe 
in \Cref{fig:obj}, 
the default orientation [see \Cref{fig:tol}(b1)] for the square grid in some
earlier work is not the best. 

A better approach (in our opinion) would be to incorporate orientation into the optimization process: we introduce a set of
binary variables $o(i,j)$, where $(i,j)\in E$ labels two qubits connected by an
edge. A value of $o(i,j)=0$ indicates \textit{control} qubit $i~\rightarrow$
\textit{target} qubit $j$, while $o(i,j)=1$ indicates \textit{control} qubit
$j~\rightarrow$ \textit{target} qubit $i$. 
These binary variables must appear in most directed constraints in the original
setting (see \Cref{table:constraint}), including
\textbf{C1}, \textbf{E1}, \textbf{E2}, \textbf{D1}, \textbf{S1}, \textbf{S1},
and \textbf{T1}. 

As an example, the previous constraint \textbf{E2}: $| f_{d,i\rightarrow j} - f_i -
\alpha| \geq \delta_{E2}$ depends on the choice of \textit{control} qubit
$\rightarrow$ \textit{target} qubit orientation. With the introduction of a
sufficiently large positive number $M$, the \textbf{E2} constraint can be
written as two constraints that allow for the orientation $o(i,j)$ to be
selected by the optimization method: 
\begin{align*}
    |f_j - f_i - \alpha| + M\cdot o(i,j) &\geq \delta_{E2}\\
    |f_i - f_j - \alpha| + M\cdot [1-o(i,j)] &\geq \delta_{E2}.
\end{align*}
Similarly, the orientation can be incorporated into other constraints, as
shown in \Cref{table:new_constraint}.
Consequently, only an undirected graph needs to be provided for optimization, with
orientation optimized during the process.

\begin{table*}[t]
\centering
\begin{tabular}{|>{\centering\arraybackslash}m{0.8cm}|>{\centering\arraybackslash}m{6.2cm}||>
{\centering\arraybackslash}m{6.2cm}|>{\centering\arraybackslash}m{2cm}|}
\hline
\textbf{Type } & \textbf{Definition $o(i,j)=0$ } &  \textbf{Definition $o(i,j)=1$ } & \textbf{Bounds} \\
\hline
\textbf{C1} & $f_i + \alpha \leq f_{d,i\rightarrow j} \leq f_i$, \hfill $(i, j) \in E$ & $f_j + \alpha \leq f_{d,j\rightarrow i} \leq f_j$, \hfill $(i, j) \in E$  & --- \\
\hline
\textbf{E1} & $\left| f_{d,i\rightarrow j} - f_i \right| \geq \delta_{E1}$, \hfill $(i, j) \in E$ & $\left| f_{d,j\rightarrow i} - f_j \right| \geq \delta_{E1}$, \hfill $(i, j) \in E$  & 17 MHz \\
\hline
\textbf{E2} & $\left| f_{d,i\rightarrow j} - f_i - \alpha \right| \geq \delta_{E2}$, \hfill $(i, j) \in E$  & $\left| f_{d,j\rightarrow i} - f_j - \alpha \right| \geq \delta_{E2}$, \hfill $(i, j) \in E$  & 30 MHz \\
\hline
\textbf{D1} & $\left| f_{d,i\rightarrow j} - f_i - \alpha/2 \right| \geq \delta_{D1}$, \hfill $(i, j) \in E$ & $\left| f_{d,j\rightarrow i} - f_j - \alpha/2 \right| \geq \delta_{D1}$, \hfill $(i, j) \in E$  & 2 MHz \\
\hline
\textbf{S1} & $\left| f_{d,i\rightarrow j} - f_k \right| \geq \delta_{S1} \hfill (i, j, k) \in N$ & $\left| f_{d,j\rightarrow i} - f_k \right| \geq \delta_{S1} \hfill (j, i, k) \in N$ & 17 MHz \\
\hline
\textbf{S2} & $\left| f_{d,i\rightarrow j} - f_k - \alpha \right| \geq \delta_{S2}$, \hfill $(i, j, k) \in N$ & $\left| f_{d,j\rightarrow i} - f_k - \alpha \right| \geq \delta_{S2}$, \hfill $(j, i, k) \in N$ & 25 MHz \\
\hline
\textbf{T1} & $\left| f_{d,i\rightarrow j} + f_k - 2f_i - \alpha \right| \geq \delta_{T1}$, \hfill $(i, j, k) \in N$ & $\left| f_{d,j\rightarrow i} + f_k - 2f_j - \alpha \right| \geq \delta_{T1}$, \hfill $(j, i, k) \in N$ & 17 MHz \\
\hline
\end{tabular}
\caption{\label{table:new_constraint} Modified constraints for undirected
graph, with $o(i,j)=0$ marks $i\rightarrow j$ and $o(i,j)=1$ marks $i\leftarrow
j$.  
The
original constraints of addressability (\textbf{A1} and \textbf{A2}) are kept.
}
\end{table*}

\subsection{Multichip Design: Periodic Boundary Condition}\label{sec:multichip}
In designing larger quantum processors, multichip designs are commonly
employed, but the impact of boundary conditions in such designs has seldom been
discussed. Analogous to applications in statistical
mechanics, boundary conditions can play a crucial role when scaling from small
systems to larger ones. Earlier
research~\cite{morvan2022optimizing,smith2022scaling} often assumed periodic
boundary conditions, which may not be optimal for SC architecture
design.

\begin{figure}[h]%
\centering
\includegraphics[width=0.99\linewidth]{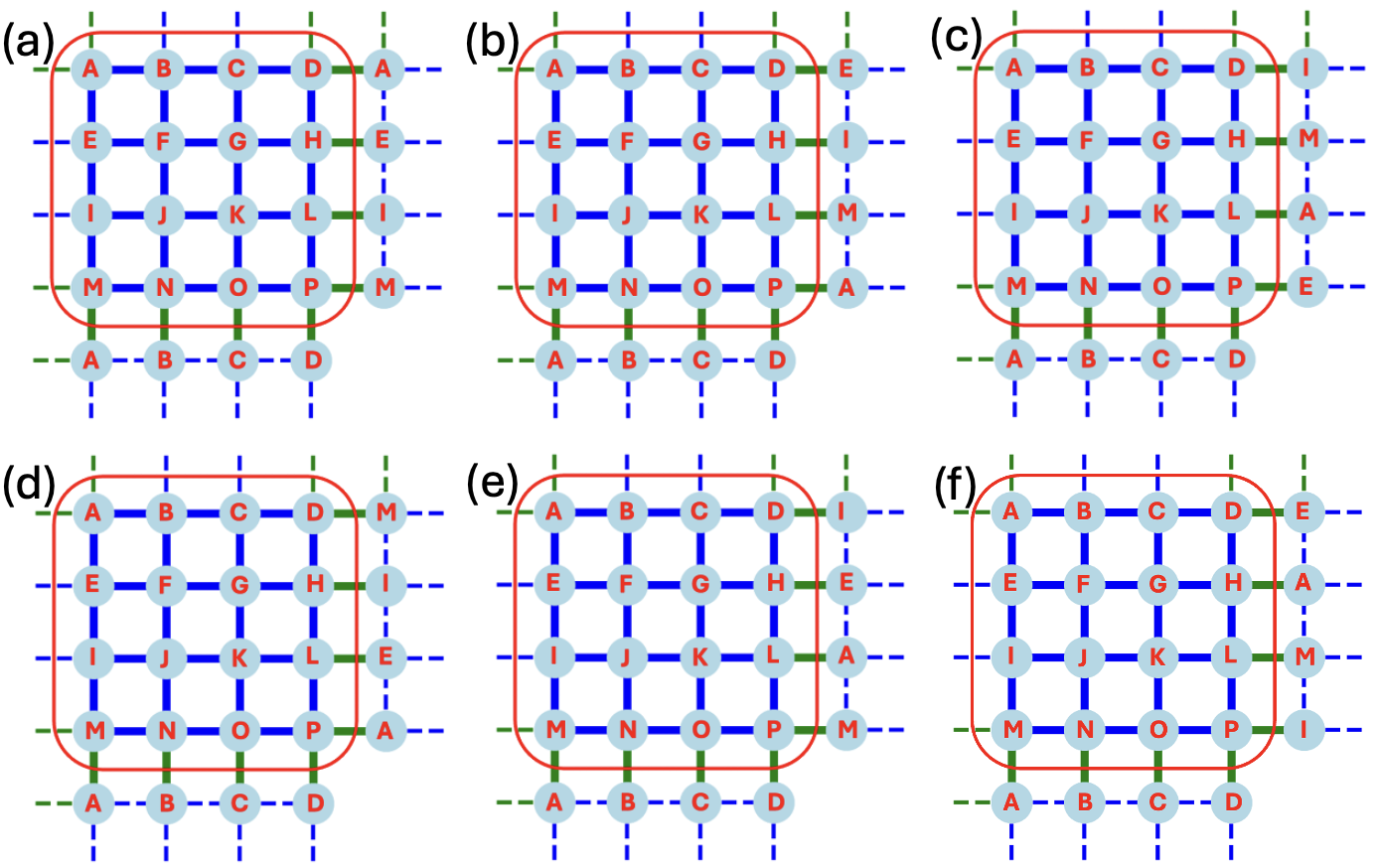}
\caption{Tested boundary conditions: (a) conventional periodic boundary condition (PBC1); (b) and (c): twisted boundary conditions (PBC2 and PBC3); (d) M\"obius boundary condition (MBC1); (e) and (f): twisted M\"obius boundary condition (MBC2 and MBC3).} \label{fig:pbc}
\end{figure}

In this study we tested various boundary conditions for multichip designs,
including twisted and M\"obius boundary conditions, as shown
in \Cref{fig:pbc}. Our results emphasize the importance of boundary conditions
in assembling multiple small chips into a larger processor. Specifically, for
assembling small square chips, our findings identify more effective boundary
conditions that generate higher yields than those previously reported.

\section{results: improved optimization technique \label{sec:results}}
We now present the effects of incorporating the new modeling approaches in
\Cref{sec:formulation}. We find that these approaches improve the solution to the frequency allocation
problems; as a result, they help  find high-yield frequency configurations for large SC processors.

\subsection{Tightening Constraints}
\label{sec:tighten}
\begin{figure}[h]%
\centering
\includegraphics[width=0.99\linewidth]{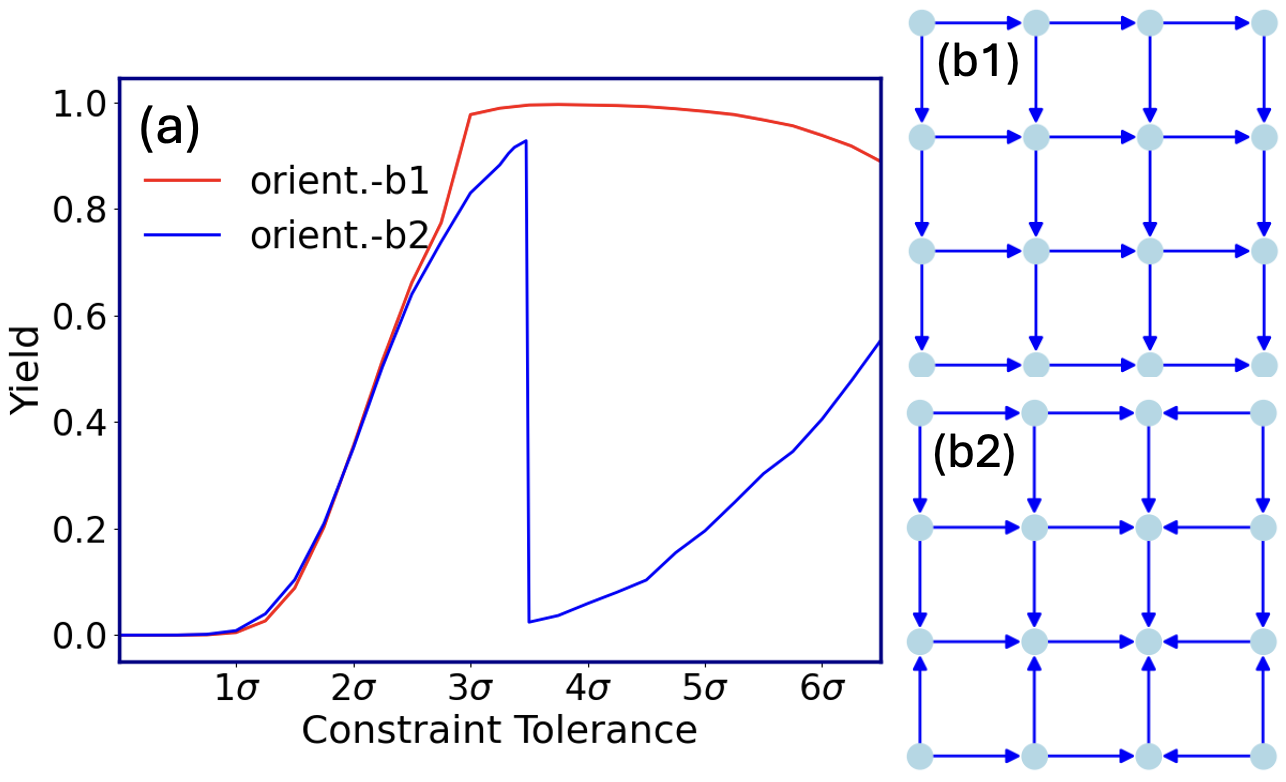}
\caption{(a) Yield at dispersion level of $\sigma$=10 MHz for two sets of
orientation, with constraint tolerance added in unit of $\sigma$, as discussed in the text. (b1) and
(b2): Preset orientations of the $4\times 4$ grid. The sudden drop in yield
observed in the (b2) curves is caused by the emergence of a different type of
constraint violation during frequency reoptimization, as discussed in the main
text.} \label{fig:tol}
\end{figure}

For a specific edge set $E$ with fixed edge orientations, some constraints can be more likely violated than the rest. 
The occurrence of these types of collisions is substantially reduced when the corresponding constraints are tightened---that is, increasing the $\bar{\delta}_n$ values before optimization. As the examples in~\Cref{fig:tol} show, properly adding tolerance
to the constraints leads to significantly higher yields. 
With (b1) orientation in~\Cref{fig:tol}, frequency collisions are likely to occur on
\textbf{A1} and \textbf{S1}; with (b2) orientation, frequency collisions are
more probable on \textbf{A2} and \textbf{S1}. To
demonstrate the improvement from constraint tightening, we added tolerance to the \textbf{A1} and \textbf{S1}
constraints for orientation (b1), and the \textbf{A2} and \textbf{S1}
constraints for orientation (b2). As shown in~\Cref{fig:tol}, by appropriately
setting these tolerance values, the yield can exceed 90\%. However, we observed
that the yield does not increase monotonically with tolerance. If the tolerance for
some constraints is too large, it can induce collisions with other constraints.
When the yield drops, conflicts on \textbf{A2}, \textbf{D1}, and \textbf{S2}  show up
in orientation (b1), and conflicts on \textbf{A1} show up in orientation (b2).
In the worst-case scenario, excessively increasing certain constraint
tolerances can eliminate the possibility of finding any feasible solutions.

These examples suggest
 maintaining a level of tightened constraints during optimization. Throughout
our work we add a tolerance of $\epsilon_\text{tol}=10$-$20$ MHz 
to all constraints
except for \textbf{D1}. The tolerance for \textbf{D1} is omitted for two
reasons: first, this constraint is rarely violated in practice; and second,
empirically, tightening this constraint sharply slows down the optimization to
find a feasible solution and eventually leads to a low-yield solution.

\subsection{Graph Orientation}
\label{sec:undirected}
\begin{figure}%
\centering
\includegraphics[width=0.7\linewidth]{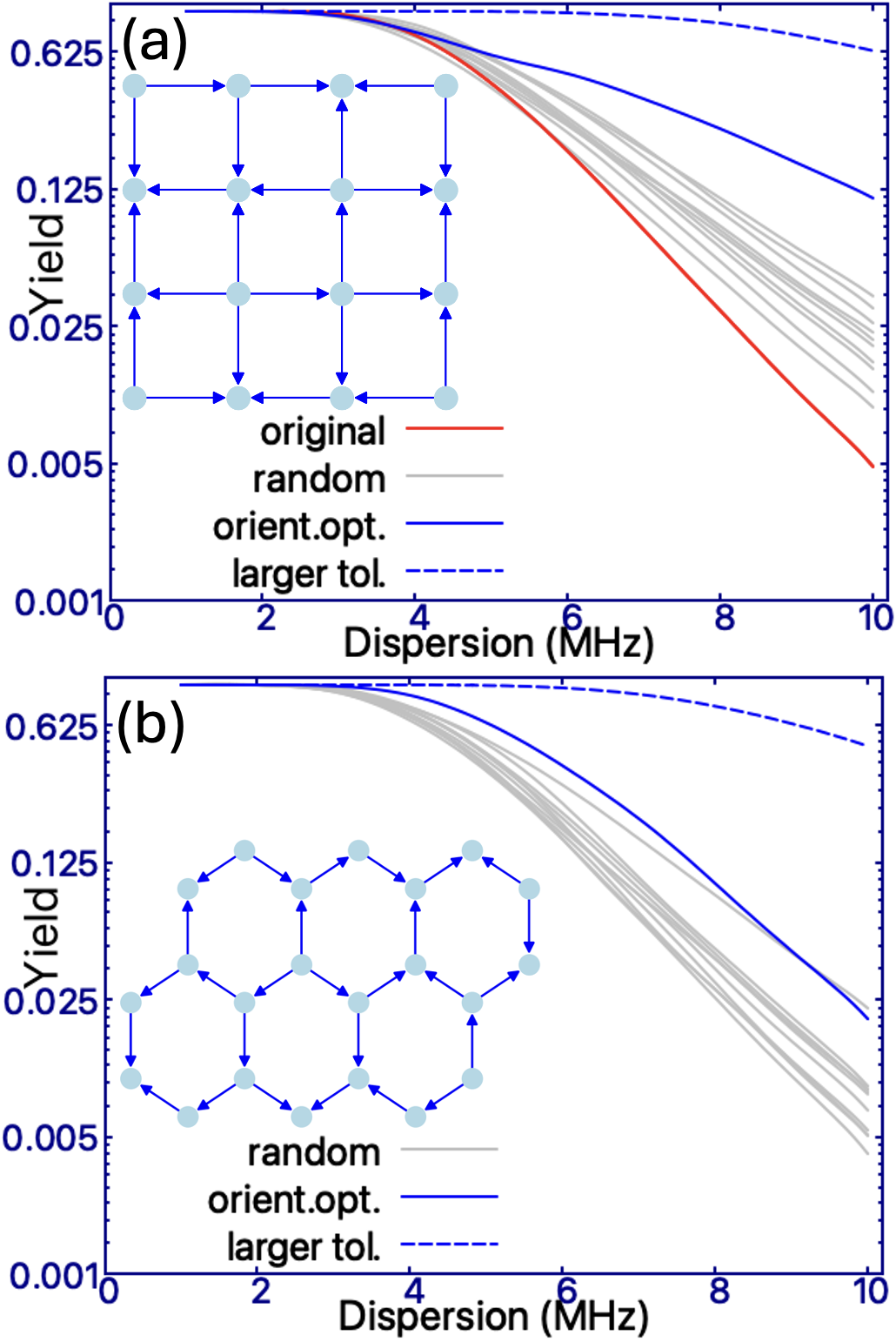}
\caption{(a) Yield at different levels of dispersion for different sets of
orientation. Red: the orientation used in Ref.~\cite{morvan2022optimizing};
grey: results from 10 random sets of orientation; blue solid: result with orientation optimization and enforced edge
differences (as in \Cref{eq:diff}). 
(b) Grey: results from 10 random sets of orientation; blue solid: result with orientation optimization and enforced edge
differences. In both graphs, the insets give
the optimized orientation. All solid curves are optimized with constraint
tolerance  of 10 MHz, while the dashed curves indicate results with orientation optimization, edge
differences imposed and larger constraint tolerance (20 MHz).}
\label{fig:direction}
\end{figure}

We now highlight how incorporating the edge orientation within the optimization process can significantly
improve yield. In \Cref{fig:direction}(a), when optimized with constraint
tolerance $\epsilon_\text{tol}=10$ MHz, the original orientation used in previous
research produces the lowest yield.
 Optimizing on the 10 random orientations as in \Cref{fig:obj} (at the same
 tolerance level), all give higher yields than
the original orientation does. At 10 MHz dispersion,
for both grid and hexagon topologies, with the same constraint tolerance,
optimizing the orientation results in over an order of magnitude increase in
yield compared with most randomly selected orientation sets. 
With greater constraint tolerance, as shown by the dashed lines in
\Cref{fig:direction}, the yield increases further.

\subsection{Local Yield for Different Boundary Conditions}
\label{sec:pbc_results}
\begin{figure}[h]%
\centering
\includegraphics[width=0.7\linewidth]{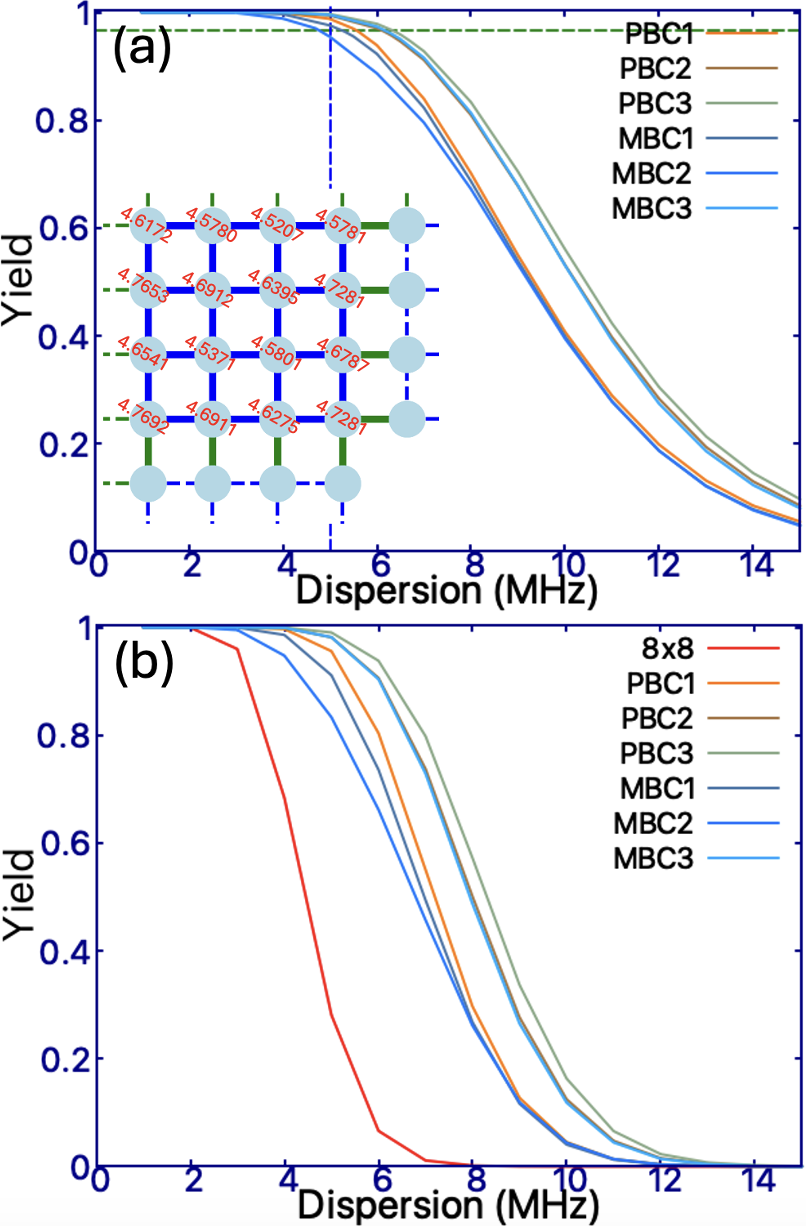}
\caption{(a) Performance of local $4\times 4$ modules: the local yield at
different levels of dispersion for different sets of periodic boundary
conditions.  Green line: effective yield of 96.5\%; Blue line: dispersion of
5 MHz. Inset: optimal frequency configuration for the best boundary condition
[PBC3 in \Cref{fig:pbc}]. (b) Assembling a $8\times 8$ chip: red: directly
optimizing a $8\times 8$ chip; others: assembling $4\times 4$ chips to a
$8\times 8$ chip.
}\label{fig:pbc_results}
\end{figure}

In assembling small modules into a large chiplet, optimized results significantly depend on boundary conditions. As an example, the frequencies of a $4\times 4$ square grid chip are optimized with the boundary conditions presented in~\Cref{fig:pbc} and a constraint tolerance $\epsilon_\text{tol}=20$ MHz; 
 the results are shown in \Cref{fig:pbc_results}(a). With the new optimization method and various boundary conditions, we achieve optimized frequency
configurations that outperform those in Ref.~\cite{morvan2022optimizing}, which
requires a frequency dispersion of no larger than 5 MHz to achieve 10\% yield for a 1,000-qubit
square grid chip. When the local yield of the $4\times 4$ module exceeds
96.5\%, combining 62 replicas of these small modules is likely to result in a
yield of over 10\%. Among
all tested boundary conditions, one of the twisted periodic boundary conditions,
PBC3, gives the best performance.

\subsection{Realizing Larger Chips}
\label{sec:assembly}
Multimodule design  not only reduces
the optimization cost of finding frequency configurations of large chips but also enhances yield under finite computing resources, provided appropriate boundary conditions are applied. \Cref{fig:pbc_results}(b) gives
an example of assembling four small modules ($4\times4$ square grids) into 
one large module ($8\times8$ square grid). These results are compared with the result of directly optimizing an $8\times8$ chip\footnote{The multimodule designs are given 20,000 CPU-seconds to
optimize with edgewise difference $\delta_\text{diff}=2$ MHz imposed, while 
directly optimizing the $8\times8$ chip is given 80,000 seconds, with edgewise difference of $\delta_\text{diff}=0.5$ MHz imposed.}. The results give clear evidence that the
multimodule design outperforms the monochip design in fixed computing
resources.

\begin{figure}[h]%
\centering
\includegraphics[width=0.99\linewidth]{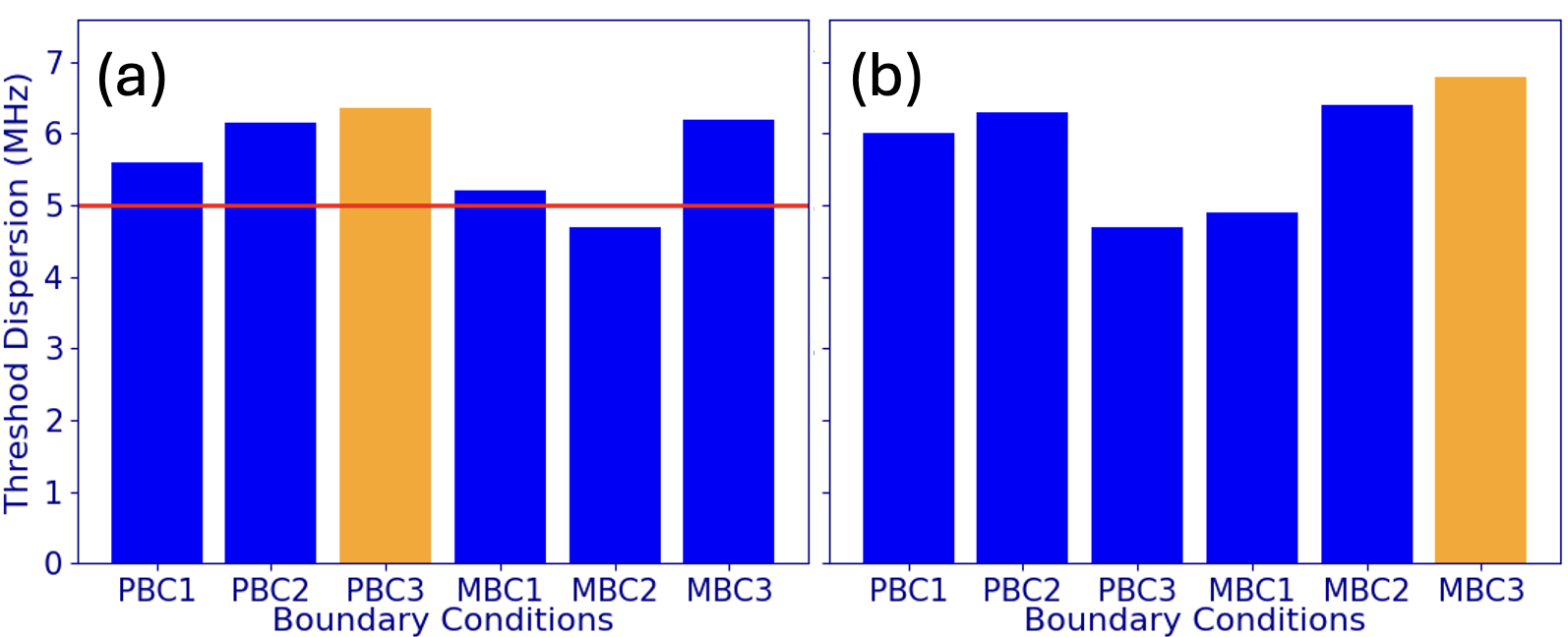}
\caption{\textbf{Multimodule assembly to 1,000 qubits}. (a) Threshold
tolerance on the dispersion of the frequency to achieve 10\% yield for a
$32\times32$ square chiplet on different boundary conditions. Red line:
dispersion of 5 MHz, which is the best from previous
research~\cite{morvan2022optimizing}. Orange bin: the best boundary condition
with the same constraint tolerance applied. 
(b) Threshold tolerance on the dispersion of the frequency to achieve a 10\%
yield for a $28\times35$ hexagon chiplet on different boundary conditions.
\label{fig:1000_qubit}} 
\end{figure}

The results also outperform results of the previous optimization method when composing a
chip with over 1,000 qubits in square grid topology. The maximum fabrication frequency variation required for a frequency configuration to achieve a 10\% yield is referred to here as the ``threshold frequency.'' When assembling 64 small
modules together into a $32\times 32$ chiplet, 5 out of 6 tested boundary
conditions give a threshold dispersion larger than that given in
Ref.~\cite{morvan2022optimizing}.
We can also see that, after optimization, some constraints are more likely
to experience collisions. As an example, for the PBC3 boundary condition, the three most
violated constraints are \textbf{A1}, \textbf{D1}, and \textbf{F1}. Further
tightening these constraints can give better results. As the optimal result we
can find, the threshold dispersion can go up to around 6.5 MHz, which is a
significant improvement from previous research. From the same method, the
hexagon grid is tested in a unit cell of $4\times5$ with different boundary
conditions\footnote{given 80,000 CPU-seconds for each run};  the results are
shown in \Cref{fig:1000_qubit}(b). The best result gives a threshold dispersion
of around 7 MHz to make a $28\times35$ hexagon chiplet, which is comparable to
the previous research.

\section{Outlook}\label{sec:outlook}
We present several improved optimization techniques to address
frequency allocation problems in SC circuit processors. These
techniques include tightening constraints, imposing edgewise differences,
optimizing graph orientations, and incorporating multimodule designs with
various boundary conditions. The results highlight the importance and
efficiency of these methods.

Specifically, the results demonstrate an increase in the dispersion threshold from 5 MHz to
approximately 6.5 MHz for manufacturing square grid processors with over 1,000
qubits. While this threshold still falls short of current transmon technology
capabilities, our findings suggest that the potential of optimization methods
for frequency allocation is yet to be fully exploited. On the other hand,
advancements in hardware, such as post-processing~\cite{gokhale2020optimized},
may help bridge the gap between this theoretical threshold and practical
fabrication limits. Future research on optimization could explore assigning
weights to objective function terms, leveraging machine learning  to link
objectives and yield, and implementing multilayered module assemblies.
Furthermore, we note that even if the optimizer does not find the
optimal allocation or if the fabricated chip does not fully meet all
constraints, a best-effort solution still holds significant value.
Specifically, techniques such as routing around dead qubits, quantum error
correction schemes, and super-stabilizer
methods~\cite{auger2017fault,lin2024codesign} can be employed to tolerate a
small number of frequency collisions.

Future work may also involve developing application-specific
designs~\cite{li2020towards,ding2020systematic} rather than general-purpose quantum processors or
optimizing total gate time that is also defined by qubit frequencies. The
optimization techniques employed in this work may also be extended to other SC
circuit devices, such as fluxonium devices~\cite{dogan2023two}, remote CR
coupling~\cite{ohfuchi2024remote}, and ZZ gates~\cite{mitchell2021hardware,huang2024fast}.
One may also expect this optimizer to contribute to tunable
couplers~\cite{hour2024context,klimov2024optimizing} as a temporally local
solver. Additionally, these methods could be applied to other quantum platforms
with qubit-level control, including Rydberg atom tweezer
arrays~\cite{radnaev2024universal} or individually-controlled trapped
ions~\cite{zhu2022multi,bond2022effect,hou2024individually}.

\begin{acknowledgments}
This work was supported by the U.S.~Department of Energy, Office of Science,
Office of Advanced Scientific Computing Research, under Contract No.~DE-AC02-06CH11357 (Accelerated Research for
Quantum Computing program, Fundamental Algorithmic Research for Quantum Utility
project) and under Award Number DE-SC0021526.
\end{acknowledgments}

\bibliography{apssamp}

~\\~\\
\framebox{\parbox{.90\linewidth}{\scriptsize The submitted manuscript has been created by
        UChicago Argonne, LLC, Operator of Argonne National Laboratory (``Argonne'').
        Argonne, a U.S.\ Department of Energy Office of Science laboratory, is operated
        under Contract No.\ DE-AC02-06CH11357.  The U.S.\ Government retains for itself,
        and others acting on its behalf, a paid-up nonexclusive, irrevocable worldwide
        license in said article to reproduce, prepare derivative works, distribute
        copies to the public, and perform publicly and display publicly, by or on
        behalf of the Government.  The Department of Energy will provide public access
        to these results of federally sponsored research in accordance with the DOE
        Public Access Plan \url{http://energy.gov/downloads/doe-public-access-plan}.}}
\end{document}